# Native defect association in $\beta$-Ga$_2$O$_3$ enables room-temperature *p*-type conductivity

Zeyu Chi[1], Corinne Sartel[1], Yunlin Zheng[2], Sushrut Modak[4], Leonid Chernyak[4], Christian M Schaefer[3], Jessica Padilla[3], Jose Santiso[3], Arie Ruzin[5], Anne-Marie Gonçalves[6], Jurgen von Bardeleben[2], Gérard Guillot[7], Yves Dumont[1], Amador Pérez-Tomás[3], Ekaterine Chikoidze*[1]

[1]Groupe d'Etude de la Matière Condensée (GEMaC), Université Paris-Saclay, UVSQ – CNRS, 45 Av. des Etats-Unis, 78035 Versailles Cedex, France

[2] Institut des Nano Sciences de Paris (INSP), CNRS UMR7588, Sorbonne Université, 4 place Jussieu, 75252 Paris, France

[3]Department of Physics, University of Central Florida, Orlando, FL 32816, USA

[4]Department of Physical Electronics, School of Electrical Engineering, Tel Aviv University, Tel Aviv 69978, Israel

[5]Institut Lavoisier de Versailles (ILV), Université Paris-Saclay, UVSQ – CNRS, 45 Av. des Etats-Unis, 78035 Versailles Cedex, France

[6]Univ. Lyon, CNRS, ECL,UCBL,INSA Lyon, CPE, Institut des Nanotechnologies de Lyon, 69621 Villeurbanne Cedex, France

[7]Catalan Institute of Nanoscience and Nanotechnology (ICN2), CSIC and the Barcelona Institute of Science and Technology, Barcelona, Spain

*ekaterine.chikoidze@uvsq.fr

**Abstract:**

The room temperature hole conductivity of the ultra-wide bandgap semiconductor $\beta$-Ga$_2$O$_3$ is a pre-requisite for developing the next-generation electronic and optoelectronic devices based on this oxide. In this work, high-quality *p*-type $\beta$-Ga$_2$O$_3$ thin films grown on *r*-plane sapphire substrate by metalorganic chemical vapor deposition (MOCVD) exhibit $\rho = 5\times10^4$ $\Omega\cdot$cm resistivity at room temperature. A low activation energy of conductivity as $E_{a2} = 170 \pm 2$ meV was determined, associated to the $V_O^{++} - V_{Ga}^-$ native acceptor defect complex. Further, taking advantage of cation (Zn) doping, the conductivity of Ga$_2$O$_3$:Zn film was remarkably increased by three orders of magnitude, showing a long-time stable room-temperature hole conductivity with the conductivity activation energy of around 86 meV. $V_O^{++} - Zn_{Ga}^-$ defect complex has been proposed as a corresponding shallow acceptor.

1. **Introduction**

Being the mainstream of the ultra-wide bandgap (UWBG) semiconductors, beta-gallium oxide ($\beta$-$Ga_2O_3$) has a bandgap ($E_g$) up to 5.0 eV, a room-temperature electron mobility ($\mu$) of higher than 200 $cm^2$/(V·s) [1,2], and a theoretically estimated critical breakdown field ($E_{CR}$) of 8 MV/cm [3], leading to high Baliga's figures of merit (BFOMs) [4]. Such properties make $\beta$-$Ga_2O_3$ attract much scientific and technological attention in optoelectronics [5,6], solar-blind photodetectors [7,8]), and power electronics (field effect transistors [3,9], Schottky barrier diodes [10,11]). However, to realize the full functionality of any emerging electronic technology based on UWBG semiconductors, both *n*- and *p*-type conductivity (i.e., bipolarity) should be attained. The room-temperature electron concentration of $\beta$-$Ga_2O_3$ can reach $10^{19} – 10^{20}$ $cm^{-3}$ by Si, Sn, Ge, and Nb doping [12–15]. While $\beta$-$Ga_2O_3$ displays significant resistance to the formation of shallow acceptor levels by doping [16,17]. In fact, the potential for achieving *p*-type or *n*-type conductivity in wide band gap materials through doping depends on the characteristics of the native point defects, such as their formation enthalpies and position relative to the valence and conduction bands. Therefore, it is critical to identify, experimentally investigate, and control the concentration of these native point defects in $\beta$-$Ga_2O_3$ through growth parameters to determine the feasibility of achieving room temperature hole conductivity.

By native point defects have been investigated theoretically, revealing that gallium oxide has a significant advantage due to the high formation energy of donor vacancies. It makes the compensation mechanism by point defects less favorable, and thus creates favorable conditions for the realization of native hole conductivity in $\beta$-$Ga_2O_3$ [18]. The origin of hole conductivity was attributed to single acceptor $V_{Ga}$ ionization energy ($E_i$) determined to be about 1.2 eV from the Hall effect measurements of the highly compensated $\beta$-$Ga_2O_3$/*c*-$Al_2O_3$ thin films grown by metal organic chemical vapour deposition (MOCVD) technique [19]. Moreover, native point defects have the ability to form complex-like defects through their associations. Indeed, the study of native vacancy pairs in $\beta$-$Ga_2O_3$ has a long history. According to Vasiltsiv *et al.* [20], the vacancies pair $V_{Ga}$-$V_O$ are charge-active centres that, together with single Ga vacancies, can act as acceptors. Binet and Gourier [21] claimed to have this native defect complex even in unintentionally doped *n*-type $Ga_2O_3$ bulk. Very recently, this defect complex was predicted to have comparable or even lower formation energy than single vacancies in Oxygen-rich conditions [22]. Furthermore, abundant gallium and oxygen vacancy complexes were identified by positron-electron annihilation signals irrespective of conductivity of bulk $Ga_2O_3$ crystals [23]. Very recently, Zhu *et al.* [24] proposed that the native $V_{Ga}$-$V_O$ divacancy defects level can be at 4.42 eV from conduction band by photoluminescence (PL), confirming the results of Binet and Gourier [21]. These complexes can play very important role in optical and electrical properties of the material. The probability of the creation of native defect associates and their impact on the electrical and optical properties of II-VI compounds has been extensively studied [25]. Several reports showed the *p*-type conductivity in ZnS can be significantly enhanced by shallow complex-like defects $V_{Zn}^{-} - V_{S}^{++}$ [26,27], or $Ag_{Zn}^{-} - V_{S}^{++}$ with Ag doping [28]. Additionally, Reynolds *et al.* [29] introduced shallow acceptor complex $V_{Zn}$-$N_O$-$H_i$ with an ionization energy of 130 meV into ZnO employing a set of growth and annealing procedures.

Can native defects in $Ga_2O_3$ exhibit similar behaviour? Interestingly, it has been observed that post-annealing in an oxygen atmosphere can result in room-temperature hole conductivity ($\rho = 1\times10^4$ $\Omega\cdot$cm) in $Ga_2O_3/c$-$Al_2O_3$, thanks to the appearance of a shallow acceptor level with 0.17 eV activation energy. This center was attributed to $V_O^{++} - V_{Ga}^{-}$ defect complex [19]. Indeed, in a close stoichiometry state, single oxygen vacancy tends to be in the double-charged ($V_O^{++}$) state rather than in the a single-charge ($V_O^{+}$) state [30]. Point defects tend to associate with one another driven by Coulomb interaction, since the system always tries to reach the state of minimum free energy. Later, the same shallow defect complex with Cathodoluminescence (CL) activation energies 88 – 101 meV, which was corroborated by electron-beam-induced current (EBIC) measurements in as grown *p*-type $Ga_2O_3/c$-$Al_2O_3$ thin films [31].

The motivation of this work was the elaboration of *β*-$Ga_2O_3$ thin films by MOCVD with important number of shallow acceptor complex defect, yielding room-temperature hole conductivity.

## 2. $Ga_2O_3$ thin films

To achieve a high number of native defects in *β*-$Ga_2O_3$ thin films, it is crucial to obtain a low compensation regime, which means minimizing the number of native donor defects. We have grown *β*-$Ga_2O_3$ by MOCVD technique on *r*-oriented sapphire substrate. It is known that *β*-$Ga_2O_3$ films grown on *r*-sapphire differs significantly from those grown on *c*-oriented sapphire substrates [32–37]. Additionnaly, Nakagomi *et al.* [38–40] reported distinct structural properties, revealed by X-ray diffraction (XRD) and transmission electron microscopy (TEM) measurements, which were attributed to the oxygen atomic arrangements on the surface of substrates. Indeed, when a *β*-$Ga_2O_3$ crystal is constructed on a sapphire substrate, gallium atoms form chemical bonds with the oxygen atoms on the sapphire surface. In the case of growth on *r*-oriented sapphire, the arrangement of three oxygen atoms and one aluminum atom in the (113) *n*-plane corresponds to the arrangement of three oxygen atoms and one gallium atom on the (-201) plane of *β*-$Ga_2O_3$ [38]. Such a well-fitted atomic arrangement is not favorable to oxygen vacancy (donor defect) creation, we can thus assume that it may be favorable for low compensation in *p*-type *β*-$Ga_2O_3$.

In our study, we have performed a detailed crystallographic study of *β*-$Ga_2O_3/r$-$Al_2O_3$ films. It was recently reported that no peak of the film is observable in *θ*/2*θ* scan and that ($\bar{2}$01) planes are tilted about 30° from the normal of the substrate [34]. Nevertheless, the authors hypothesized that OP peaks should be too weak to measure and concluded that the plane of the films should be (100) and/or (001). Thus, there are ambiguities about *β*-$Ga_2O_3$ films on *r*-sapphire. We propose here to deepen the structural analysis.

We got the same results in the *θ*/2*θ* scan for undoped *β*-$Ga_2O_3$ films grown on *r*-sapphire as in the work cited [34], *i.e.* no peak except 012 of the substrates and its harmonics. In *β*-$Ga_2O_3$ $\bar{2}$01 PF (**Figure 1 (a)**), two spots of $\bar{2}$01, labelled *A* and *A'*, were indeed found out of the normal of the film by *χ* = ± 30.2° toward $Al_2O_3$ *a*-axis azimuth. Orientations *A* and *A'* are equivalent by mirror symmetry of the substrate. Two other spots labelled *B* and *B'* were found at the high tilt angle of *χ* = 80.8°. Orientations *A* and *B* are not equivalent with respect to the substrate. The intensity of spot *B* is almost two orders of magnitude lower than that of spot *A*, indicating that the proportion of orientations *B* and *B'* is less than 5% of the film. In *β*-Ga$_2$O$_3$ 001 PF (**Figure 1 (b)**), four spots were also observed. It can be noticed that no spot is located at the centre of the PF, *i.e.* (001) is not the film plane. Knowing that the angle between the spots $\bar{2}01$ and 001 should be about 49.9°, each spot of 001 can be coupled to a spot of $\bar{2}01$ and therefore labelled as orientations *A*, *A'*, *B* or *B'*.

The angular positions of the spots $\bar{2}01$ and 001 fully fix orientations *A* and *B* of the film and allow to deduce all the other orientations. We still cross-checked and measured the PFs of 111 and 400 reflections. For the 400 PF with the diffraction angle of 30.06°, the reflections $\bar{4}01$ and 110 with closing diffraction angles of respectively 30.49° and 30.32° were also observed. Reflections collected and related respectively to orientation *A* and to orientation *B* are summarized in **Figure 2**. Orientation *A* can be reproduced from the epitaxy of *β*-Ga$_2$O$_3$($\bar{2}01$) on *c*-sapphire (sample C) which was grown simultaneously with the one on *c*-sapphire (sample R), by a rotation of 152° around the normal to the substrate and then a tilt of -30.2° toward the substrate *a*-axis. The angular positions of the spots measured agree with those calculated with the bulk lattice parameters, indicating that the film is unstrained within the measurement accuracy.

For orientation *B*, some spots such as 400 and 110 are common to orientation *A* but should be labelled self-consistently with reversed indices. It follows that this is the case for all hk0 reflections. Inverting hk0 reflections between orientations *A* and *B* corresponds to a rotation of 180° around the *c*-axis of the unit-cell, the *c*-axis being unchanged and common to the two orientations. In other terms, orientation *A* and *B* can be understood as twins.

Strain in the film is detailed by XRD and GIXRD scans (**Figure 3**). As there is no strictly OP reflection for the present case, we considered $\bar{4}01$ reflection of orientation *A* (denoted $\bar{4}01_A$) and 401 reflection of orientation *B* (denoted $401_B$) which are off the sample normal by about 14° (**Figure 2**). These reflections provide then almost OP strain of orientations *A* and *B*. Their measurements were done in asymmetric geometry by *ω/2θ* and *ω* scans (**Figure 3 (a), (c)**). Similarly, $3\bar{1}4_A$ reflection is off the plane by only 0.8° and $5\bar{1}\bar{4}_B$ reflection by 0.5°. The reflections provide information on IP strain and were measured by IP scans (**Figure 3 (b), (d)**). The measured Bragg angles correspond to those expected with an uncertainty of 0.1 – 0.2%, in agreement with the PF result that strain is not sizable within the measurement accuracy. The absence of noticeable strain also results in a small misorientation of the planes (Insets in **Figure 3**).

In the case of growth on *r*-sapphire substrate, the plane of the Ga$_2$O$_3$ films does not correspond to any basic lattice plane. Two main equivalent orientations have ($\bar{2}01$) planes tilted respectively ±30.2° toward the substrate *a*-axis. Two minor equivalent orientations constitute less than 5% of the film and can be seen as twins of the main orientations. No strain was measured in the film. The OP and IP misorientation of the planes is small and within ±0.2°.

It remains that the epitaxy of the film on *r*-sapphire is intriguing and leaves questions that should be explored further. If the plane of the film does not correspond to any basic plane, what is the exact film/substrate interface to achieve epitaxy? One possibility would be that lattice matching would occur from facets that could be present on the surface of the substrate. We found by X-ray reflectivity a very high surface roughness (> 5 nm) in the sample. It indicates that the growth does not occur layer-by-layer but rather in a 3D fashion. The growth mode and the presence of minor orientations could also promote strain relaxation in films on *r*-sapphire.

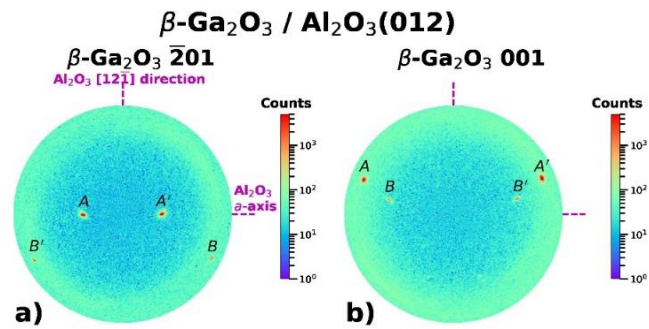

**Figure 1.** Pole figures in sample R. (a) $\beta$-Ga$_2$O$_3$ $\bar{2}$01 pole figure where the azimuths of Al2O3 a-axis and [12$\bar{1}$] directions are indicated by magenta dashed lines. A and A', as well as B and B', mark a pair of equivalent spots by mirror symmetry of the substrate. (b) $\beta$-Ga$_2$O$_3$ 001 pole figure.

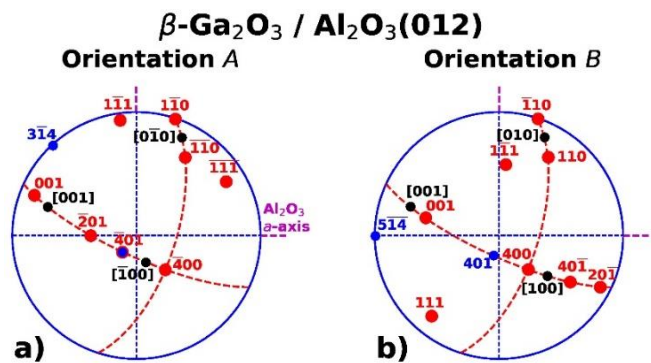

**Figure 2.** Orientation projections of reflections measured in sample R and related to orientation A (a) and to orientation B (b). The reflections collected in PF measurements are indicated by red dots. The reflections used in IP and asymmetric OP scans are symbolized by blue dots. The orientations of the unit-cells deduced, namely that of the invariant c-axis ([001]), are indicated by black dots. The red dashed circles outline the sets of hk0 and h0l reflections.

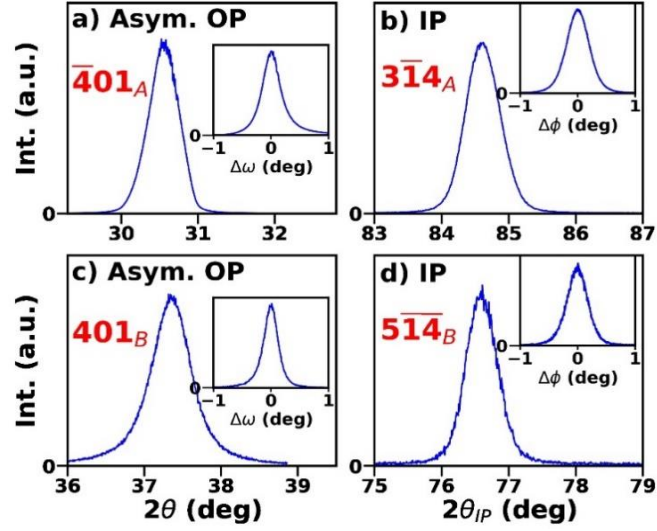

**Figure 3.** XRD and GIXRD scans in sample R. Reflections are subscripted by A and B for orientations A and B, respectively. Orientation A is characterized by asymmetric OP $\bar{4}01_A$ reflection (a) and quasi-IP $3\bar{1}4_A$ reflection (b). Orientation B is characterized by asymmetric OP $401_B$ reflection (c) and quasi-IP $5\bar{1}\bar{4}_B$ reflection (d). OP rocking curves ($\omega$-scans) and corresponding IP azimuthal curves ($\phi$-scans) are in the insets. $2\theta_{IP}$ is the IP diffraction angle measuring the IP component of the diffracting vector.

**Figure 4 (a)** shows the top view scanning electron microscopy (SEM) image of the $Ga_2O_3$/$r$-$Al_2O_3$ sample, it exhibits a coarser morphology with grains look like needles (ridges). The thickness was determined at 453 nm (**Figure 4 (b)**). **Figure 4 (c)** represents the Raman scattering spectra of the $r$-plane sapphire substrate and the $Ga_2O_3$/$r$-$Al_2O_3$ sample. Observed spectral features of $Ga_2O_3$ can be unambiguously attributed to the $\beta$-phase phonon lines [41–43]. In addition to XRD results, it evidences that the sample has the single $\beta$-phase. The spectrum is comparable with our previously reported Raman spectrum of $\beta$-$Ga_2O_3$ on the $c$-sapphire substrate [19]. Reflectance and transmittance measurements on undoped $Ga_2O_3$/$r$-sapphire were carried out at room temperature in 200 – 2500 nm wavelength range, having high transparency over 80% at visible and near-infrared wavelength ranges, as shown in **Figure 4 (d)**. The absorption coefficient $\alpha$ was then calculated, $(\alpha h\upsilon)^2$ vs. $h\upsilon$ (Tauc plot) was used to estimate the optical bandgap. There is only one contribution to the bandgap at $4.60 \pm 0.04$ eV as shown in the inset of **Figure 4 (d)**.

The electrical resistivity for undoped $Ga_2O_3$/$r$-$Al_2O_3$ was measured from 300 to 850 K in a Van der Pauw configuration. Four electrical contacts of silver paste were painted at each corner of the square-shaped (1×1 cm$^2$) sample and then annealed at high temperature. **Figure 4 (e)** shows the temperature dependent resistivity, with $\rho = 5.4 \times 10^4$ Ω·cm resistivity at 300 K. Two slopes can be distinguished from the $\ln(\sigma)$ versus $1/T$ plot, the activation energy of conductivity is determined: $E_{a1} = 414 \pm 7$ meV and $E_{a2} = 170 \pm 2$ meV (**Figure 4 (f)**). $E_{a1}$ most probably corresponds to single $V_{Ga}$ related defect acceptor center when sample is of low compensation regime, while the second $E_{a2}$ to $V_O^{++} - V_{Ga}^-$ native defect association. [19] From the room temperature Hall effect measurements, the hole concentration and mobility are $p = 5.7 \times 10^{13}$ cm$^{-3}$ and $\mu < 1$ cm$^2$/(V·s), respectively.

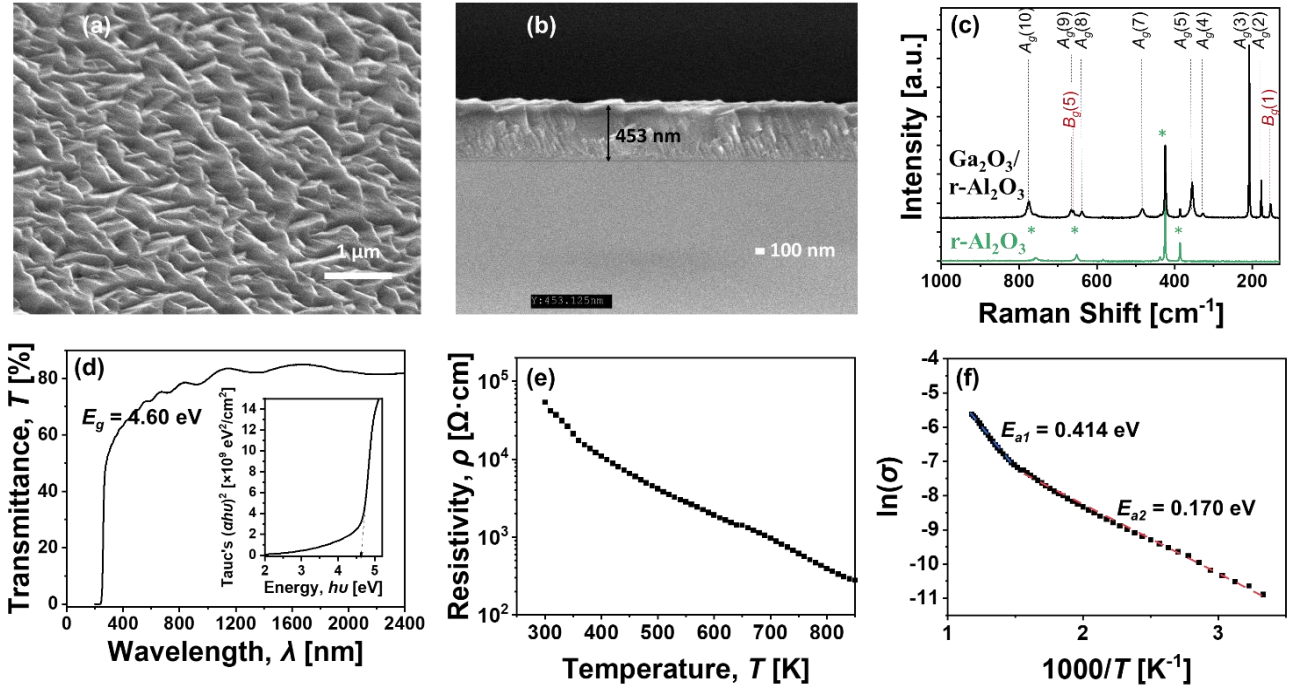

**Figure 4.** (a) Top view SEM image of β-Ga$_2$O$_3$/r-Al$_2$O$_3$ sample. (b) SEM cross-section image showing the thickness of 453 nm for the thin film. (c) Raman spectra of r-plane sapphire substrate (phonon modes labelled "*") and the Ga$_2$O$_3$/r-Al$_2$O$_3$ sample. (d) Room temperature transmittance spectrum for Ga$_2$O$_3$/r-Al$_2$O$_3$ sample, the optical bandgap ($E_g$) = 4.60 eV estimated by Tauc's plot (inset). (e) Temperature dependent resistivity in the range of 300 – 850 K, (f) Two activation energies were determined as $E_{a1}$ = 0.414 eV and $E_{a2}$ = 0.170 eV by the ln($\sigma$) versus 1/$T$ plot for undoped β-Ga$_2$O$_3$/r-Al$_2$O$_3$.

This result is significant as it suggests that β-Ga$_2$O$_3$, similarly to wide band gap II-VI semiconductors, tends to create native defect associations that act as shallow acceptor centres. It should be underlined that evidently, the probability of defects pairing strongly depends on the concentration of native defects ($V_O$ ; $V_{Ga}$) and the vicinity of their positions. These favorable conditions are determined by the growth parameters (temperature, pressure) and the crystal structure.

The possibility of obtaining a room temperature hole conductivity in undoped MOCVD Ga$_2$O$_3$ with an activation energy of $E_{a2}$ = 170 meV encouraged us to attempt to further increase the conductivity. To this end, we need to make shallower the $V_O^{++} - V_{Ga}^{-}$ complex. To this end, we opted for the cation doping.

### 3. Enhancement of hole conductivity by Zn doping

With our extensive experience in using Zn as a dopant, we have opted to incorporate it into β-Ga$_2$O$_3$ to increase the hole conductivity. It was reported on the base of electron magnetic resonance (EPR) that, Zn in tetrahedral and octahedral sites acts a deep acceptor (0.65 eV and 0.78 eV, respectively) [44]. Consistent with this finding, we demonstrated that at the doping level of 10$^{16}$ cm$^{-3}$, Zn is indeed a deep acceptor in β-Ga$_2$O$_3$ thin films with an ionization energy of 0.7 eV [45]. Computational study suggested that substitutional Zn dopant atoms may replace Ga atoms on the tetrahedral $Ga_{(1)}$ sites, which were predicted to be preferential sites [46,47]. However, it is noted that hybrid functional calculations and experiment found that Zn atoms may substitute both tetrahedral and octahedral sites with nearly equal energies [44,48].

Upon replacing a Ga atom, Zn reduces the oxygen vacancy concentration decreases, promoting the recharging of $V_O^+$ and placement of $V_O^{++}$ in the nearest neighbor site to Zn, and play donor's role for positive charge ($Zn_{Ga}$ defect) compensating [49]. Our thermodynamic analysis [50] has shown that as the concentration of incorporated Zn increases (from $10^{15}$ to $10^{18}$ cm$^{-3}$), the concentration of double-charged oxygen vacancy $V_O^{++}$ also significantly increases (up to $10^{18}$ cm$^{-3}$). This will strongly promote the association of two native acceptor-donor defects $V_O^{++}$-$Zn_{Ga}^-$, since the proximity of $V_O^{++}$ to $Zn_{Ga}^-$ can reduce its electronic level.

**Figure 5** illustrates the $V_O^{++} - Zn_{Ga}^-$ complex-like defects in Zn doped $\beta$-Ga$_2$O$_3$ Further experimental studies, such as high resolution transmission electron microscopy, infra-red spectroscopy, positron-electron annihilation, are required to identify the exact sites and nature of defect complexes in Zn:$\beta$-Ga$_2$O$_3$.

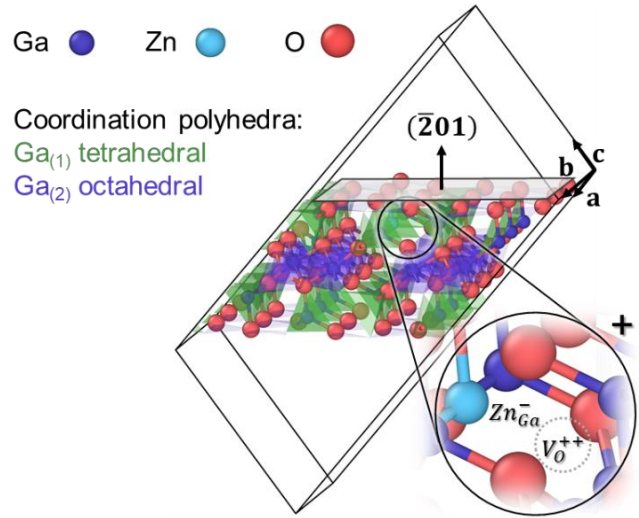

**Figure 5.** Hypothetical crystal structure model of Zn-doped $\beta$-Ga$_2$O$_3$ containing a $V_O^{++} - Zn_{Ga(1)}^-$ defect complex. A slab cut between ($\bar{2}$01) planes within a 2×3×2 replicated unit cell is depicted. The polyhedral highlight tetrahedral (green) and octahedral (purple) coordination of the $Ga_{(1)}$ and $Ga_{(2)}$ atoms by O atoms, respectively. A substitutional Zn atom replacing Ga on a tetrahedral $Ga_{(1)}$ site is highlighted. OVITO Pro 3.7.12 was used for visualization [51] based on a $\beta$-Ga$_2$O$_3$ crystal structure file (Data retrieved from the Materials Project for Ga$_2$O$_3$ (mp-886) from database version v2022.10.28 [52])

The Zn doped $\beta$-Ga$_2$O$_3$ thin film was grown on *r*-plane of 1-inch sapphire substrates. Secondary-ion mass spectroscopy (SIMS) analysis showed that Zn concentrations in the samples are [Zn] = 1.0 ×10$^{18}$ cm$^{-3}$. XRD analysis does not display any significant difference in structure and strain for Zn:$\beta$-Ga$_2$O$_3$ thin film with respect to undoped $\beta$-Ga$_2$O$_3$ thin film, nor the surface morphology by SEM images, while the thickness was determined at 320 nm. The optical bandgap at room temperature for all Zn doped $\beta$-Ga$_2$O$_3$ film is around $E_g$ = 4.74 eV ± 0.05 eV, which is slightly larger than for undoped samples. All bandgaps in doped and undoped films were determined by the same Tauc plot.

## A. Electronic structure and Valence band of Zn:Ga$_2$O$_3$

X-ray photoelectron spectroscopy (XPS) has been employed to study the above-mentioned Zn incorporation effect. The room temperature photoemission spectrum for the Ga2$p$, Zn2$p$, and O1$s$ core levels is shown in **Figure 6 (a)-(d)**. The binding energy value of the core level for the Ga2$p_{3/2}$ states at 1119.3 eV (as determined by Michling *et al*. [53]) on the cleaved $\beta$-Ga$_2$O$_3$ single crystals was used for calibrating the binding energy positions of the XPS spectra. The XPS spectra from the surface of a control sample of a 500 nm thick commercial (Novel Crystal Technology, Inc.) epilayer of nominally *n*-type Si-doped $\beta$-Ga$_2$O$_3$ (N$_d$ - N$_a$ = 1.3×10$^{18}$ cm$^{-3}$) grown on a single crystal $\beta$-Ga$_2$O$_3$ were used as a benchmark for the $\beta$-Ga$_2$O$_3$/*r*-sapphire core levels and valence band determination. The XPS spectrum has again been calibrated with respect to the Ga2$p$ energy at 1119.3 eV. The Zn:Ga$_2$O$_3$ Ga2$p$ doublet separation between the primary Ga2$p_{3/2}$ and the secondary Ga2$p_{1/2}$ was determined to be $\Delta$Ga2$p$ = 26.8 eV, while the peak of Ga2$p_{1/2}$ appeared at 1146.12 eV **(Figure 6 (a))**. For the reference Si:Ga$_2$O$_3$ sample, those values were determined to be very similar with $\Delta$Ga2$p$=26.9 eV and Ga2$p_{1/2}$ at 1146.22 eV. For both, the Ga2p states show a reasonably sharp structure with a full width at half maximum (FWHM) of ~1.8 eV. As shown in **Figure 6 (b)**, there is a small asymmetric contribution of Ga-Zn states appearing at larger binding energies in agreement with our previous article [54]. The Zn2$p$ split-orbit component doublet was determined to be at 1023.32 eV and 1046.62 eV, for Zn2$p_{3/2}$ and Zn2$p_{1/2}$, respectively and with a separation of $\Delta$Zn2$p$ = 23.3 eV shown in **Figure 6 (c)**. The value of the Ga2$p_{3/2}$ is larger than the usual value reported for ZnO (~1021 – 1023 eV) [55] which again is believed to be due to the contribution of Zn-Ga states. The main O1s level for Zn:Ga$_2$O$_3$ appeared at 532.22 eV while the reference Si:Ga$_2$O$_3$ was 532.12 eV **(Figure 6 (d))**. The O1$s$ level exhibited secondary (O$_{II}$) and tertiary (O$_{III}$) shoulders at ~533.5 eV and ~534.9 eV or some ~1.3 eV and ~2.9 eV from the main Ga-O peak, respectively. This triple contribution shape of the O1$s$ core level was similar to the one determined for the reference and was ascribed to ad-atoms at the surface as hydroxides (-OH) and C=O, for (O$_{II}$) and (O$_{III}$), respectively [56]. There are, however, some O1$s$ features visible when compared to the reference sample. There is a small Zn-O contribution at low binding energies (~530 eV) compatible with Zn-O bonds. Besides, the Zn:Ga$_2$O$_3$ O1$s$ peak was found to be wider and shifted towards higher binding energies (compared to the Si:Ga$_2$O$_3$ reference), and an indication of potential intermixing of Ga-O (and Zn-O) oxidation valence.

It has now been well established that the monoclinic gallium oxide valence band density of states is predominantly of O2$p$ character (as it usual in metal oxides) while the conduction band states are predominantly derived from Ga4$s$ and Ga4$p$ states mixed with some contributions of O2$p$. In **Figure 7 (a)-(b)**, valence band photoemission spectra from our layers are shown. The three maxima of the O2$p$ valence band correspond to the three different oxygen sites in $\beta$-Ga$_2$O$_3$ (O$^{(1)}$, O$^{(2)}$ and O$^{(3)}$). The valence band width for both, the reference Si-doped sample and the Zn-doped sample, is around 8 eV, which agrees with previous reports [57]. Nevertheless, an exceptionally large number of tail states extend deeper in the bandgap for the $\beta$-Ga$_2$O$_3$:Zn film surface compared to the reference *n*-type $\beta$-Ga$_2$O$_3$ sample. This, in practice, indicates a much-reduced Fermi level value (from the edge of the valence band), with the equilibrium energy level located in the lower part of the bandgap and an acceptor *p*-type nature of the layer surface. The presence of such states related to divacancy complex close to the valence band edge have already been investigated in previous ab-initio studies such as the one of Usseinov *et al.* [22]. As shown in **Figure 7 (d)**, the experimental relative density and position of the divacancy states, compared to the O2p states, also are also in agreement with these atomistic simulations. Among all combinations of paired vacancies, the creation of $V_{Ga1}$-$V_{O(1,2,3)}$ complexes is energetically more favorable than similar $V_{Ga2}$-$V_{O(1,2,3)}$ complexes and the lowest-energy configuration is for the $V_{Ga1}$-$V_{O1}$ pair vacancy.

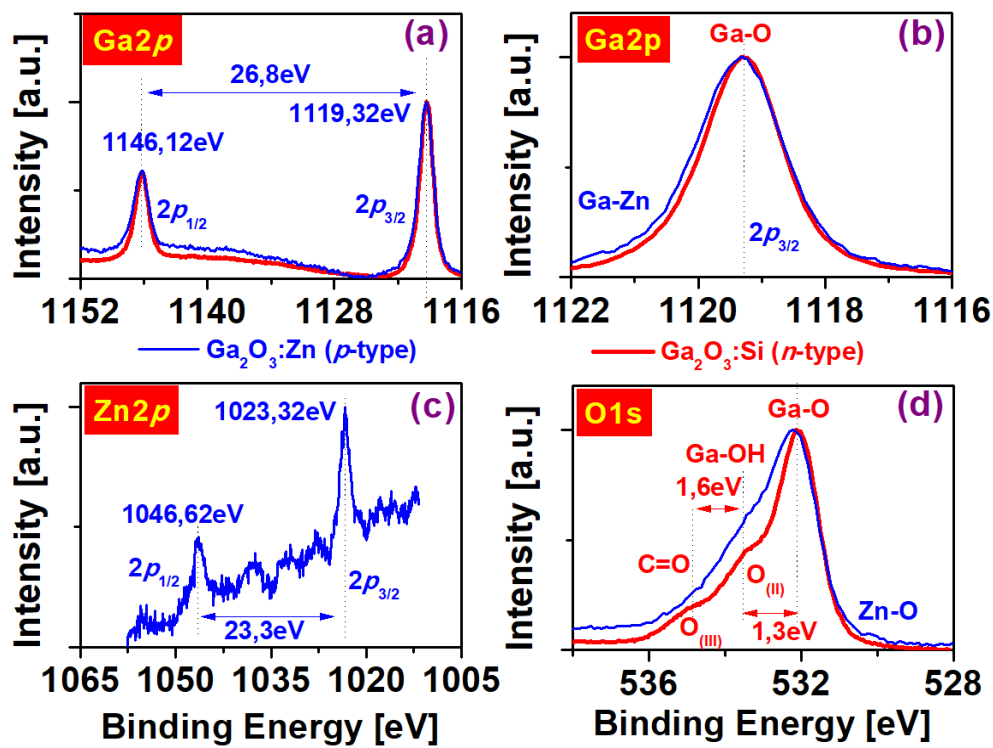

**Figure 6.** Detail (intensity normalized) of the main XPS peaks for (a) gallium (Ga2$p_{1/2}$ – Ga2$p_{3/2}$), (b) gallium (Ga2$p_{3/2}$), (c) Zinc (Zn2$p$), (d) oxygen (O1$s$), of Zn doped Ga$_2$O$_3$ grown on *r*-Al$_2$O$_3$ substrate.

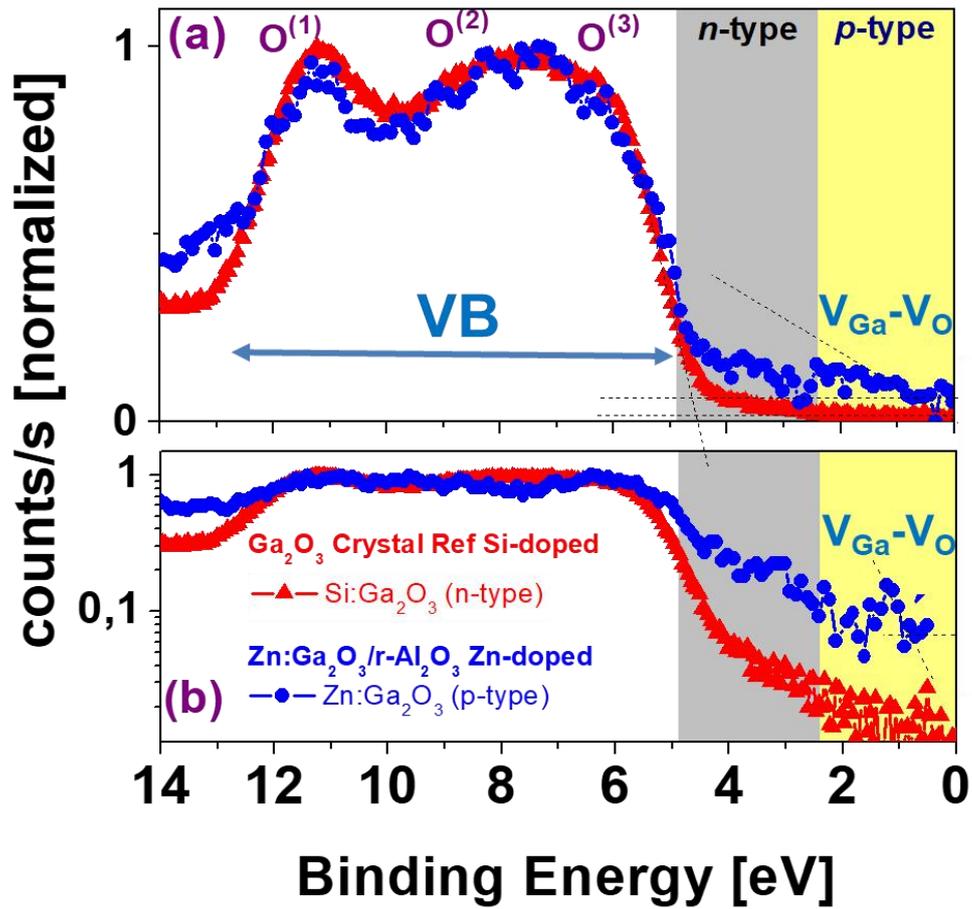

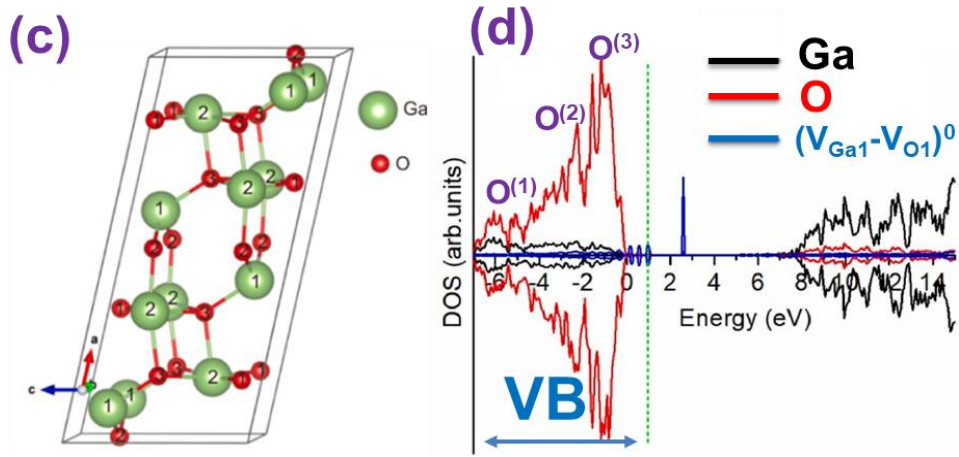

**Figure 7.** (a) Experimental XPS high-resolution valence band of Zn doped $\beta$-Ga$_2$O$_3$ thin film, and Si doped $\beta$-Ga$_2$O$_3$ single crystal as a reference. (b) Idem (log-scale), for evidencing the tail states signature of acceptors at the valence band region (VBM). (c) Unit cell of $\beta$-Ga$_2$O$_3$ with unique positions of Ga and O denoted (Ga$_{(1)}$, Ga$_{(2)}$; O$_{(1)}$, O$_{(2)}$, O$_{(3)}$). (d) Total density of states for $V^0_{Ga(1)}$, $V^0_{O(1)}$ and $(V_{Ga(1)} - V_{O(1)})^0$ (defects states in band gap region have been magnified by a factor 5, green-dashed-vertical line represents the Fermi level) [22].

### B. Optical properties of Zn:Ga$_2$O$_3$

Photoluminescence spectra (at room temperature) of undoped and Zn-doped Ga$_2$O$_3$ sample exhibit a broad emission band between 350 – 620nm and an additional band at around 330nm for only undoped sample (**Figure 8 (a)**). The later ultraviolet luminescence (UVL) band is commonly ascribed to the recombination of self-trapped excitons, considering the absence of near-band edge emission and their lack of *β*-$Ga_2O_3$ for sub-bandgap excitation [31]. As for the broadband, it is very similar to what has been reported for our undoped $Ga_2O_3$ thin films with four characteristics luminescence bands (after deconvolution): ultraviolet (UVL' and UVL) at 375 nm and 415 nm; blue luminescence (BL) at 450 nm; and green luminescence (GL) at 520 nm.

We have studied Zn doped sample in details by cathodoluminescence measurements at (20 – 140 °C) temperatures. We observe thermal quenching of CL signal as the temperature is raised from 20 to 140 °C. Raw CL spectra for these two temperatures are shown in **Figure 8 (b)**. The plot of integrated CL intensity as a function of inverse temperature is shown in **Figure 8 (c)**, which illustrates the thermal quenching of the CL emission with increase in temperature. This behavior is empirically modelled with the following equation:

$$I(T) = I_0/\left(1 + e^{\Delta E_{CL}/kT}\right) \quad (1)$$

Here, *I(T)* is the temperature dependent integrated CL intensity, $I_0$ is a scaling constant, $\Delta E_{CL}$ is the temperature dependent activation energy pertaining to the thermal quenching effect, *k* is the Boltzmann constant and *T* is the temperature in Kelvin. The activation energy $\Delta E_{CL}$ obtained from the fit to equation 1 was 548 meV. The thermal quenching of CL emission with increase in temperature can be explained with de-trapping of electrons from meta-stable trap levels. This behavior is also seen in other wide bandgap semiconductors such as GaN [58–60] and ZnO [61]. Additionally, thermal quenching of blue (emission peak centred at 2.85 eV) photoluminescence within the temperature range of 300 – 550 K was observed by Binet and Gourier [21]. According authors a fast electron–hole recombination occurs at the acceptor site with strong electron–phonon coupling giving rise to the blue emission. An estimated activation energy of 0.42 eV in *n*-type $Ga_2O_3$ bulk crystal ($E_g$ = 4.9 eV) related to this emission is close to $\Delta E_{CL}$ observed in this report. This activation energy was attributed to hole de-trapping or the native acceptor related ionization energy. For our sample we can deduce that there is acceptor center (Ga vacancy related) located at around 0.55 ± 0.04 eV from the top of the valence band.

The obtained activation energy is for the aggregate CL intensity for the limited temperature range of 20 °C to 140 °C. Expanding the temperature range down to 4 K on the lower side and Gaussian decomposition of the constituent participating defect levels would reveal a complete picture of the CL probing of carrier recombination dynamics; This aspect will be explored in detail in a future study.

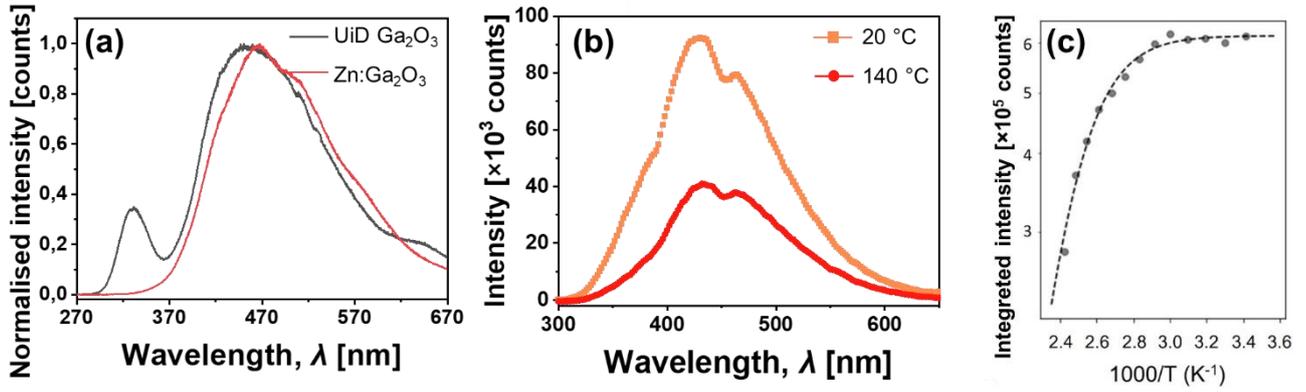

**Figure 8.** (a) Photoluminescence spectra for undoped (UID in the figure) and Zn doped samples at 294K. (b) Comparison of cathodoluminescence emission spectra for Zn doped $Ga_2O_3$ on *r*-sapphire at 20 and 140 °C. (b) Thermal quenching of CL observed in Zn doped $Ga_2O_3$ on *r*-sapphire in the temperature range of 20 to 140 °C. The activation energy obtained from equation 1 is 548 ± 42 meV.

### C. Electrical transport properties of Zn:$Ga_2O_3$

Since $Ga_2O_3$ is well known for possibility of exhibiting the surface conductivity [57,62–65], for precaution we have conducted study to eliminate this hypotheses in our case. We performed an electrochemical investigation using cyclic voltammetry. The surface states of the sample implanted in the electrolyte can be analyzed based on information such as anodic and cathodic significant currents and potentials, the oxidation and reduction onset potentials obtained by the voltammograms [66]. Indeed, various semiconductor materials such as InP [67], $TiO_2$ [68], 3C-SiC [69], GaN [70] have been studied by voltammetric techniques. **Figure 9 (a)** shows the first cycle of the voltammogram performed on the Zn doped $Ga_2O_3$ thin film. The current density was calculated by the current divided by the area immersed in the electrolyte, it is 0.5 $cm^2$ for this sample. Although the potential was swept from -3.3 to 2.7 V, i.e. the polarization is larger than the Zn:$Ga_2O_3$ energy band gap, neither anodic nor cathodic significant current, which is characteristic of a surface 2D conduction, was observed. Knowing that the electrochemical response of oxidation or reduction is caused by the mobility of the free carriers (loss or gain depending on the polarity) on the interface sample/electrolyte. Observing such a low current without significant electrochemical response reveals that the room temperature conductivity of the Zn doped $Ga_2O_3$ sample does not originate from the surface. (Moreover, the curve is not perfectly flat, possibly due to the resistance of the electrical contact between the metal and the sample).

The same electrical transport measurement configurations were used for undoped $β$-$Ga_2O_3$ and Zn doped $β$-$Ga_2O_3$/*r*-$Al_2O_3$ films. **Figure 9 (b)** shows the linear current−voltage (I−V) behavior at 300 K, indicating the Ohmic characteristics of the contacts. The first measurement showed that this sample exhibits a room temperature $ρ = 4.3×10^1$ Ω·cm resistivity. The sample was remeasured at room temperature after 18 months, resulting $ρ = 6.1×10^1$ Ω·cm. By Seebeck effect measurements, the positive sign of the Seebeck coefficient (*S*) was determined, $S = + (7 – 10)$ μV/°C at 100 – 160 °C, indicating clearly that the majority of carriers are holes. Obtaining such a significantly stable in time hole conductivity $ρ = 2×10^{-2}$ $Ω^{-1}·cm^{-1}$ at room temperature, is very important. The next step was to determine activation energy of conductivity. Resistivity versus temperature curve suggests that the sample exhibits a semiconducting behaviour. (**Figure 9 (c)**). Measurement has been carried out down to liquid nitrogen temperature, showing $\rho = 1.6 \times 10^5$ $\Omega \cdot$cm at 80 K. Plotting the ln($\sigma$) versus $1/T$ plot reveals a single slope indicating the only one conduction mechanism with activation energy determined as $E_a = 86 \pm 0.4$ meV (**Figure 9 (d)**).

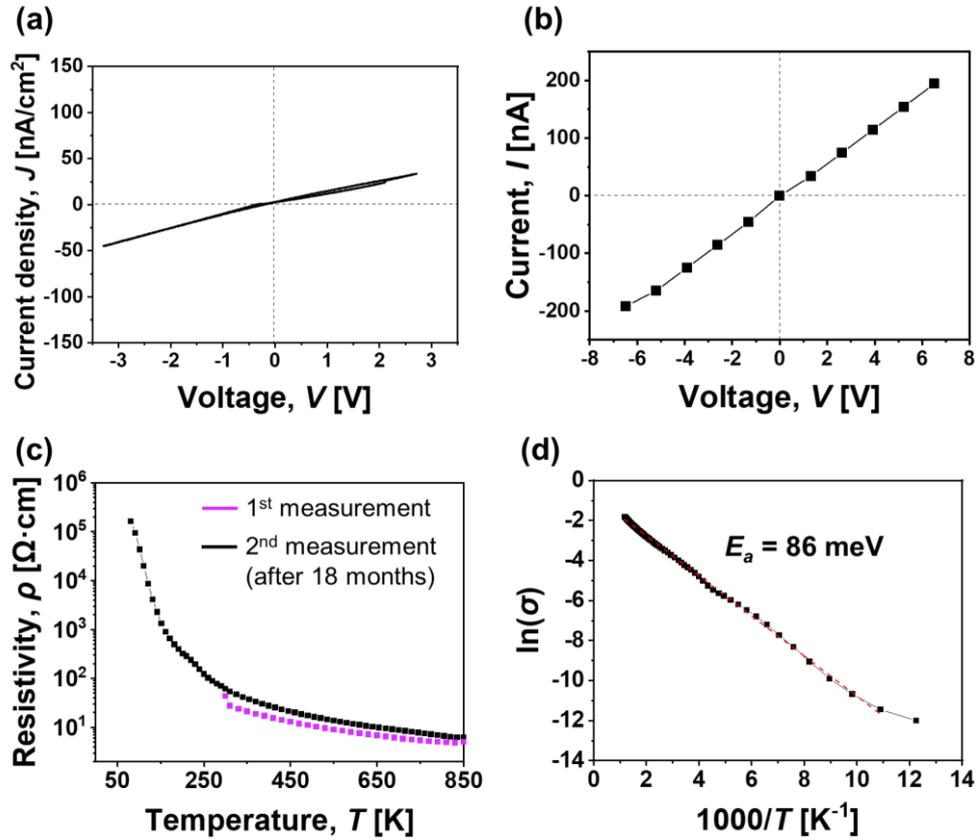

**Figure 9.** (a) Ohmic I−V characteristics of electrical contacts at 300 K. (b) Cyclic voltammogram for Zn:Ga$_2$O$_3$ thin film. (c) Temperature dependent resistivity in the range of 80 − 850 K. (d) The activation energy of the conductivity was determined as $86 \pm 0.4$ meV by the ln($\sigma$)versus $1/T$ plot in the temperature range of 80 − 850 K.

It is clear from experiment that Zn doping can decrease activation energy as compared to the undoped sample. To estimate the carrier concentration at room temperature, Hall effect measurement was carried out with a particular procedure to eliminate error with a 15-second delay between current applying and voltage measurements) at an applied magnetic field of 1.6 T. Consequently, the hole concentration and the hole mobility at room temperature were determined as $p = 1 \times 10^{17}$ cm$^{-3}$ and $\mu < 1$ cm$^2$/(V$\cdot$S), respectively. Low hole mobilities might be attributed to strong scattering processes of carriers on a significant number of native defects centres and large effective masse of holes.

## 4. Conclusion

The hole conductivity of the promising UWBG oxide *β*-Ga$_2$O$_3$ has been a hot issue recently, because it is a pre-requisite for the development of the *β*-Ga$_2$O$_3$ based next-generation bipolar electronic and optoelectronic devices. Generally, the possibility to obtain hole conductivity strongly depends what kind of point defects are dominant in the material (Frenkel, Schottky or associates), which itself depends on growth parameters of thin films. It is known that in wide band gap II-VI compounds realization of hole conductivity with doping is mainly related not with dopant related acceptor defect, but with creation of native defects charge states to assure electroneutrality of the system [25,71]. The motivation of our work was to demonstrate that in *β*-Ga$_2$O$_3$ as a ultrawide band gap III-VI material can obey the same rule, i.e., native defects can enable the realization of high hole conductivity thanks to shallow acceptor centre.

We demonstrated in this work that Zn doping of *β*-Ga$_2$O$_3$/*r*-sapphire thin film grown by MOCVD technique can exhibit a long-time stable room-temperature hole conductivity with the conductivity activation energy of around 86 meV. The origin of this level might be attributed then the donor-acceptor complex $V_O^{++} - Zn_{Ga}^{-}$. We believe that this study will add new evidences and will help to break a "taboo" related to the feasibility of room temperature hole conductivity in Ga$_2$O$_3$ via traditional growth technique and doping. We showed that the effectiveness of acceptor dopant depends on the concentration and the energetic level of cation and anion vacancies. We hope that we can inspire researchers to further study experimentally the point defects in *β*-Ga$_2$O$_3$. From the practical point of view, we expect our investigation will increase the portfolio of techniques to achieve the much wanted homoepitaxial *p-n* junction for the emerging Ga$_2$O$_3$ ultra-wide semiconductor power electronics technology enabling, for example, proper ultra-high power PiN diodes.

## 5. Experimental details

**Thin film growth:** The undoped and Zn doped $Ga_2O_3$ samples were grown in a RF-heated horizontal MOCVD reactor on *r*-sapphire ($Al_2O_3$) substrates. During the growth of undoped $Ga_2O_3$ samples, the flow rate of trimethylgallium (TMGa), and oxygen were kept at 12 µmol/min and 1200 sccm, respectively. The growth temperature, pressure and time were set at 775 °C and 30 torr, 85 minutes, respectively. For the Zn doped $Ga_2O_3$ sample, keeping the same growth conditions and the precursors' flow of Ga and $O_2$, the Zn precursor diethylzinc (DEZn) flow was set at 2.3 µmol/min, the growth time was 60 mins. The TMGa and DEZn bubbler temperatures were fixed at -10 °C and 0 °C, respectively. The thickness of undoped and Zn doped $Ga_2O_3$ were determined by MEB to be 453 and 320 nm, respectively.

**Raman spectra:** Raman spectra Raman spectra were recorded using a Renishaw Invia Reflex micro-Raman spectrometer at room temperature. The samples were excited using a cw Modu-Laser Stellar-REN laser emitting at 514.5 nm with a power of 2 – 4 mW. The reflecting microscope objective was 50 with a NA 0.75; the excitation spot diameter was 1 mm. The back-scattered light was dispersed by a monochromator with a spectral resolution of 1.4 $cm^{-1}$. The light was detected by a charge coupled device. The typical accumulation time was 30 s. Raman shifts were calibrated using an optical phonon frequency (520.5 $cm^{-1}$) of a silicon nanocrystal. Several reference Raman spectra for beta-Ga2O3 single crystal substrates and sapphire c-cut substrates have been also measured.

**Optical transmittance/reflectance measurements:** Optical transmission and reflectance spectra were measured in the 200 − 2500 nm wavelength range with a step of 2 nm using a Perklin Elmer LAMBDA 950 spectrophotometer.

**X-ray diffraction measurements:** The structure of the samples was studied using X-ray diffraction (XRD). Data were collected on a 5-circle diffractometer (Rigaku SmartLab) with Cu $K_α$ radiation from a rotating anode. The film out-of-plane (OP) orientation and strain were measured by *θ/2θ* scans and rocking curves (*ω*-scans). When no strictly OP reflection exists for a film, reflections close to the normal to the sample were measured in asymmetric OP geometry by ω/2θ scans. For OP scans, Cu $K_{α1}$ radiation was selected by a channel-cut Ge(220) 2-reflection monochromator. The film in-plane (IP) orientation and strain were probed by grazing incidence XRD (GIXRD) at the constant incidence of 0.5° without the monochromator. When no strictly IP reflection exists for a film, reflections close to IP were measured. The film global orientation texture was explored by x-ray pole figures (PFs), using the IP PF configuration [72].

**Secondary-ion mass spectrometry (SIMS) measurements:** SIMS measurements were performed using a Cameca IMS 7f equipment onto the samples to access the concentration and depth distribution of Zn dopant.

**X-ray photoemission spectroscopy (XPS) measurements:** XPS were performed with a Phoibos 150 analyzer (SPECS GmbH, Berlin, Germany) in ultra-high vacuum conditions (base pressure $3×10^{-10}$ mbar). XPS measurements were performed with a monochromatic Al Ka X-ray source (1486.74 eV).

**Cathodoluminescence measurements:** Cathodoluminescence (CL) spectra were measured insidea Phillips XL30 scanning electron microscope (SEM) with Gatan MonoCL2 attachment. The temperature of the specimen was varied with an integrated temperature-controlled stage in the temperature range of 20 °C to 140 °C. The CL signal is collected with the help of a parabolic mirror assembly and recorded with the help of a monochromator (grating blazed at 1200 lines/mm) and a Hamamatsu photomultiplier tube with a range of 150 – 850 nm.

**Hall effect measurements:** Ohmic contacts were prepared by silver paint at the four corners of the sample. Hall Effect measurements were performed by home built high impedance high temperature setup in a Van der Pauw configuration in the temperature range of 300 – 850 K applying the magnetic fields perpendicularly to the film plane varying from 0 to 1.6 T. All the measurements were carried out under

**Electrochemical measurement:** Electrochemical experiments were performed in dark using the three electrodes set-up PARSTAT 2273 potentiostat/galvanostat.Pt wire and 1 mol dm$^{-3}$ KCl silver/silver Chloride (Ag/AgCl, E°s = 199 mV vs. SHE) electrode respectively served as counter and reference electrodes. Zn doped $Ga_2O_3$ thin film was used as working electrode.

**Acknowledgement:**

The present work is a part of "GALLIA" International Research Project, CNRS, France. GEMaC colleagues acknowledge financial support of French National Agency of Research (ANR), project "GOPOWER", CE-50 N0015-01.

Research at the University of Central Florida and Tel Aviv University was partially supported by the US-Israel Binational Science Foundation (award 2018010) and NATO (SPS MYP G5748).

Researchers would like to acknowledge Dr. François Jomard (GEMaC) for SIMS measurement and analysis.